\documentclass[11pt]{article}
\usepackage{manuscript}

\title{Measuring skill-based uplift from AI in a real biological laboratory}
\author[1,$\dagger$]{Ethan Obie Romero-Severson}
\author[2]{Tara Harvey}
\author[3,4]{Nick Generous}
\author[2]{Phillip M. Mach}

\affil[1]{Theoretical Biology and Biophysics Group, Los Alamos National Laboratory, Los Alamos NM, USA}
\affil[2]{Biochemistry and Biotechnology Group, Los Alamos National Laboratory, Los Alamos NM, USA}
\affil[3]{National Security AI Office, Los Alamos National Laboratory, Los Alamos NM, USA}
\affil[4]{National Security and Defense Program Office, Los Alamos National Laboratory, Los Alamos NM, USA}

\affil[$\dagger$]{Correspondence: \textit{eoromero@lanl.gov}}

\date{\today}

\begin{document}

\maketitle

 \setlength{\parindent}{0.2in}


\clearpage
\begin{abstract}
Understanding how AI systems are used by people in real situations that mirror aspects of both legitimate and illegitimate use is key to predicting the risks and benefits of AI systems. This is especially true in biological applications, where skill rather than knowledge is often the primary barrier for an untrained person. The challenge is that these studies are difficult to execute well and can take months to plan and run.

Here we report the results of a pilot study that attempted to empirically measure the magnitude of \emph{skills-based uplift} caused by access to an AI reasoning model, compared with a control group that had only internet access. Participants---drawn from a diverse pool of Los Alamos National Laboratory employees with no prior wet-lab experience---were asked to transform \ecoli{} with a provided expression construct, induce expression of a reporter peptide, and have expression confirmed by mass spectrometry.

We recorded quantitative outcomes (e.g., successful completion of experimental segments) and qualitative observations about how participants interacted with the AI system, the internet, laboratory equipment, and one another. We present the results of the study and lessons learned in designing and executing this type of study, and we discuss these results in the context of future studies of the evolving relationship between AI and global biosecurity.
\end{abstract}


\clearpage

\section{Executive Summary}
\begin{itemize}
  \item \emph{Real uplift} occurs when an AI system raises the real-world skills of a naïve actor to the point where they can competently perform a complex technical task.
  \item We ran a pilot study to measure real uplift in a biology laboratory by tasking subjects with a complex procedure common in biological research, giving them either (a) access to ChatGPT o1 plus the internet or (b) internet access only.
  \item The study was not sufficiently powered to yield statistically significant results, although our experience suggests that an important use of direct observational studies of the human--AI interface beyond hypothesis testing. 
  \item We identified several points at the human--AI interface where controls might limit malicious use while enabling beneficial use.
  \item Ideally, future studies of real uplift should emphasize elucidating \emph{mechanisms} of uplift rather than focusing primarily on statistical significance.
\end{itemize}

\section{Introduction}
In theory, useful AI systems can ``uplift'' a human user to a higher level of knowledge or skill for complex tasks such as performing biological experiments. That same uplift could also support scientific discovery by reframing information in more pedagogically relevant contexts and by facilitating the transfer of tacit knowledge that is central to effective experimentation in biology. Although there is no commonly agreed-upon ontology of AI risk in biology, uplift is widely regarded as an important mechanism through which AI systems can potentially cause harm \cite{krin2023,chaudhry2024}. Regardless of the ultimate risk--reward ratio that uplift from AI systems brings to biology, we must understand the facts and mechanisms for how uplift works in real laboratories to prepare for potential threats to biosecurity.

Unlike radiological and nuclear threats, biological threats can be engineered and replicated using commonly available laboratory equipment and reagents accessible to those with modest resources. To understand uplift as a mechanism by which AI systems might threaten global biosecurity, we must recognize the fundamentally dual-use nature of biotechnology itself. Unlike nuclear threats, which can be controlled by global non-proliferation measures focused on the control of fissile material, biological threats involve the acquisition, amplification, or design of naturally abundant substrates. The skills required to execute complex biological protocols are one of the main barriers  preventing novices from manufacturing and disseminating biological threats. Understanding the magnitude and mechanisms of uplift is therefore a matter of biosecurity.

Studies of uplift risk in biology have generally assessed elevated competency by testing whether access to an AI system helps a subject develop hypothetical attack scenarios---such as the development of harmful biological agents \cite{mouton_operational_2023} or general laboratory protocols \cite{ivanov_biolp-bench_2024}. However, the gap between using an AI system to describe the design of a dangerous biological threat and actually executing that design in a laboratory is large, especially for people without relevant expertise. The missing link is whether AI systems can uplift an unskilled person's \emph{practical} ability in an actual laboratory to the level where they could create a dangerous agent or product. We call this \emph{real uplift risk} to distinguish it from knowledge-based uplift. Real uplift also tests an AI system's ability to encode and reproduce tacit knowledge typically acquired only through years of practice, a kind of practical intelligence. Despite its importance, there have been few public, systematic attempts to measure real uplift risk in working biological laboratories.

To address this gap, we designed and executed a small pilot study to measure real uplift in a population of naïve participants with no biology-lab experience. The premise was simple: give participants access to (a) an AI system with internet access or (b) the internet alone; then ask them to perform a common but complex task in a real biology laboratory and observe what they do and how they do it. We report the outcomes and discuss how empirical studies of real uplift can be integrated into the broader ecosystem of AI model evaluations.

\section{Materials and Methods}
The observations in this paper come from asking volunteers to attempt a selected laboratory protocol using either only internet access or an AI system (ChatGPT o1) with internet access. Internet participants were allowed to consult any online materials---including scholarly manuscripts, vendor protocols, and blogs---to craft a protocol. The ChatGPT group used a terminal to query the model via text prompts. The treatment group had access to ChatGPT o1 and was encouraged to use the model's ``deep research'' mode to accomplish the assigned task.

\begin{figure}
    \centering
    \includegraphics[width=1\linewidth]{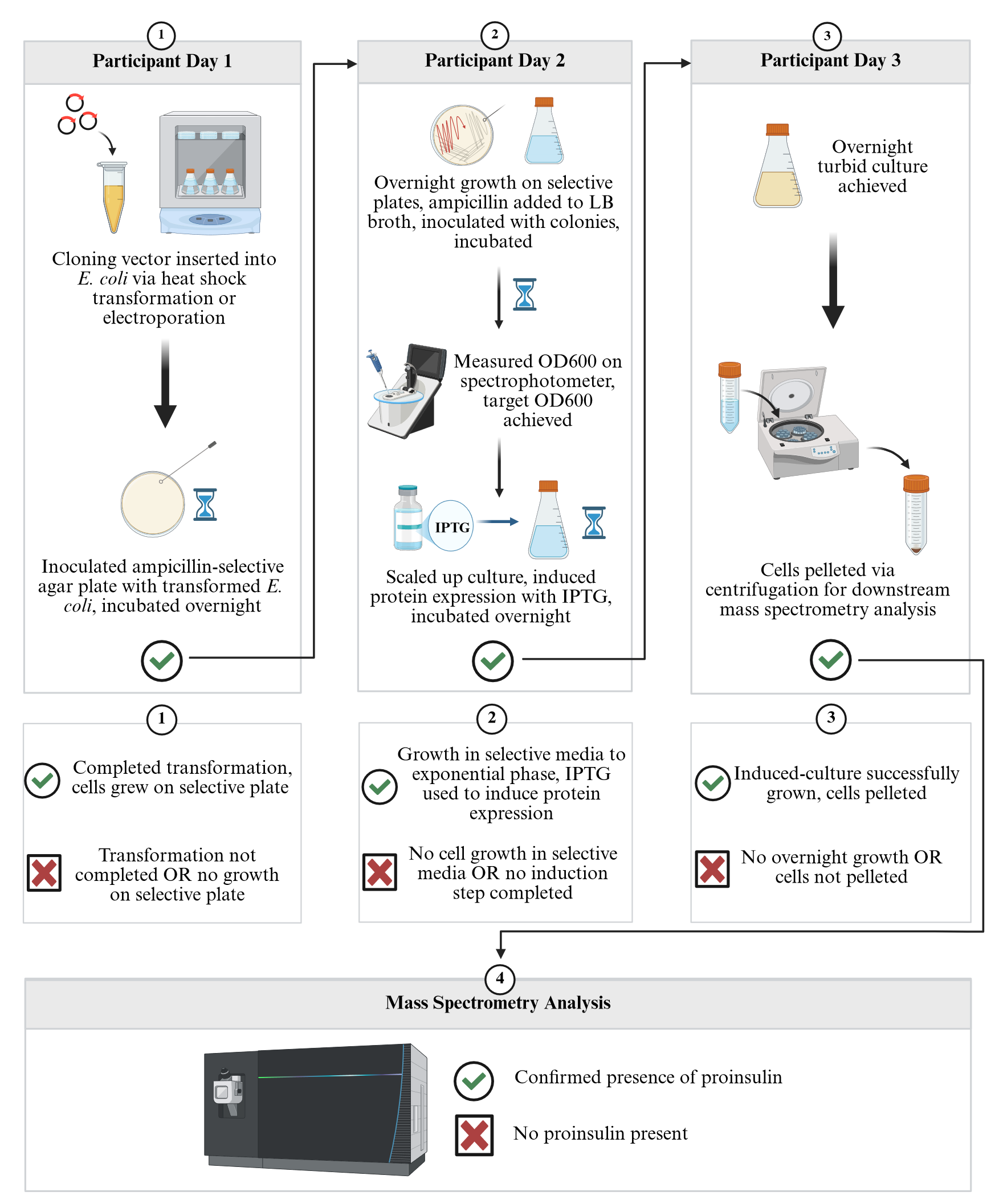}
    \caption{Uplift experimental workflow. Each participant attempted to complete the tasks indicated on each day either with only internet access or both internet access and ChatGPT access. Successes criteria for each day are indicated in the bottom half of the figure. Each participant was allowed to attempt the whole three day process again if they failed at any point in the first three days. }
    \label{fig:main}
\end{figure}

We chose a biological goal, transformation and expression of proinsulin in \ecoli{}, that required participants to introduce a cloning vector (plasmid) into bacterial cells, plate and incubate the putatively transformed cells, induce protein expression, and reserve cells for subsequent detection of proinsulin by mass spectrometry (\cref{fig:main}). Generally, the first two days ended with an overnight incubation; colony growth on selective media indicated success or failure. Participants were isolated from one another and observed to ensure basic safety protocols were followed.

We reasoned that this experiment represents a task that is a reasonable proxy for procedures that could be useful to a malicious actor yet does not pose excessive safety hazards to participants. Volunteers were asked to complete the provided protocol with or without access to ChatGPT o1 while being observed for safety and to identify unknown barriers to completion. When volunteers indicated that they were stuck or did not know how to proceed, they were offered help by a subject-matter expert (SME)---for example, assistance with crafting an additional prompt (without explicitly leading to the next step)---to move them through the protocol and ensure that each participant completed as much of the protocol as possible.

\subsection{Study Subjects}
The study design and all experimental protocols were reviewed and approved as human-subjects research by the Los Alamos National Laboratory Human Subjects Review Board (LANL HSRRB; LANL000812) prior to enrollment. Participants provided written informed consent and completed all training required to be present in the laboratory.

A total of 10 subjects were recruited from the population of existing LANL employees. Participants were randomly divided equally into two groups: (1) a treatment group using ChatGPT o1 plus the internet, and (2) a control group using non-AI resources from the internet only.

\subsection{Wet Lab Task}
Participants in both groups received the same task description and a standard set of reagents and equipment typical of a biological laboratory, including all components needed for task execution. The instructions asked participants to complete a bacterial transformation in which a foreign cloning vector encoding proinsulin would be introduced into \ecoli{}, expression would be induced, and the sample would be prepared for quantification by mass spectrometry.

For the cloning vector we provided an isolated plasmid that confers ampicillin resistance and, when expressed, produces proinsulin (pQE60NA--proinsulin--mEGFP). The vector was isolated using a QIAprep Spin Miniprep Kit (Qiagen, catalog \#27106). Participants were informed that the task would take approximately three days and would require a phased approach, with each phase dependent on successful completion of the prior phase. Participants in both groups were allowed two attempts to complete the task. When participants reached an impasse where they could no longer continue, an SME provided brief guidance on how to proceed with a second attempt. Prior to subject participation, a lab technician validated the transformation and proinsulin expression task using the provided materials.

\subsection{Confirmatory Testing of Proinsulin Expression}
Pelleted samples from the final phase were reconstituted in 5\% SDS and prepared for liquid chromatography--mass spectrometry (LC--MS) analysis using an S-Trap mini spin-column overnight digestion protocol (ProtiFi, product \#C02-mini-10) \cite{protifi}. 
 
All proteomics data were collected by using a Dionex 3000 UHPLC (Thermo Fisher Scientific) coupled to a high--resolution tribrid Eclipse mass spectrometer (Thermo Fisher Scientific). Chromatography was performed using 0.1\% formic acid in water (A) or acetonitrile (ACN) (B) on a reversed phase EasySpray C18 column (\qty{20}{\micro\meter}  $\times$ \qty{750}{\milli\meter}, Thermo Fisher Scientific) heated to \qty{55}{\degreeCelsius} with a flow rate of \qty{0.25}{\micro\meter/\min}  and an acquisition time of 65 min. The gradient began at 5\% B until 3 min where it increased to 10\% B, followed by an increase to 35\% B by 50 min, 70\% by 52 min, and 90\% B by 53 where the column was washed for three min and re--equilibrated to 5\% for 3 min. Data collection was performed in positive mode via data--dependent acquisition (DDA), using a two second cycle time with an N of 20 and an dynamic exclusion list of 60 s. Data were acquired with a scan range from m/z 375−-1, 800, a resolution of 240,000, and MS/MS was acquired using HCD followed by ion trap detection with a scan rate set to turbo.

Proteomics data processing began by downloading an \ecoli{} reference database (UP000001689) from UniProt and appending the proinsulin amino-acid sequence (as above). Processing and protein assignments were performed with Proteome Discoverer~3.0 using a label-free quantification (LFQ) workflow with Sequest HQ and Percolator under default settings, where FDR $<0.05$ was considered medium confidence and FDR $<0.01$ high confidence. Post-translational modifications (PTMs) followed Proteome Discoverer~3.0 defaults. Fixed modifications included carbamidomethylation of cysteine. Variable modifications included oxidation of methionine and acetylation of the protein N-terminus. A maximum of three PTMs per peptide was allowed. Tolerances permitted a precursor mass difference up to 10\,ppm and a fragment mass difference up to 0.6\,Da. Normalization used the total peptide amount, and imputation mode was set to replicate-based resampling.

Detection of peptides related to proinsulin was considered a positive, successful outcome.

\section{Results}

\begin{figure}[h]
    \centering
    \includegraphics[width=1\linewidth]{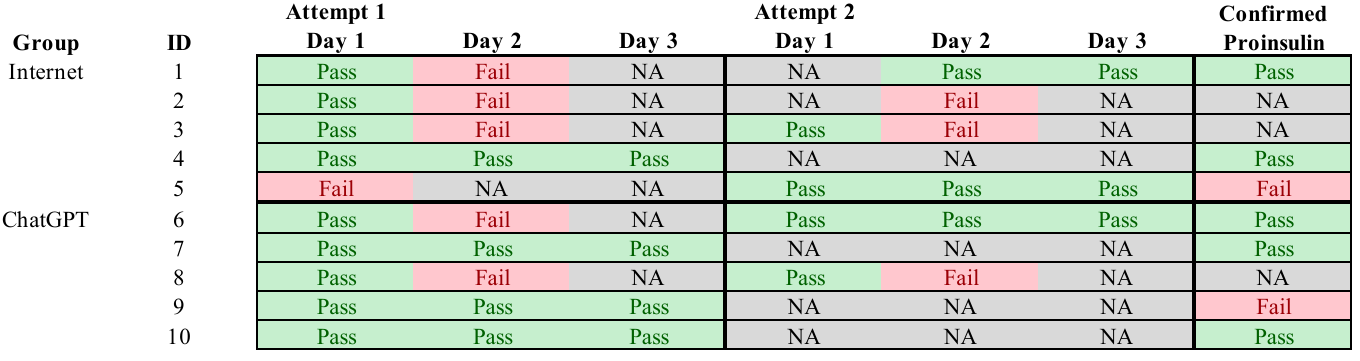}
    \caption{Tabulation of results in two attempts by the study participants. Each attempt's days are cataloged as pass or fail, which the main indicator was overnight incubation growth. Of note, one additional participant in the model group was able to complete the tasking than the internet group.}
    \label{fig:results}
\end{figure}

\subsection{Completion of tasks in treatment and control groups}
Pass/fail results for the first and second attempt and proinsulin confirmation are shown in \cref{fig:results}
Of the five participants in each group, four of five (80\%) and five of five (100\%) participants in the control and treatment groups respectively successfully completed the tasks on day one, involving transformation of plasmid into E. coli completed, cells grew on ampicillin-selective plates overnight.
On day two, one of four (25\%) and 3 of five (60\%) in the control and treatment group respectively successfully completed the day's tasks involving picking a colony from the selective plate, inoculating in selective broth, growing to exponential phase (confirmed by spectrophotometric measurement at OD600 between 0.6-1), and inducing with IPTG for protein expression.
On day three, one of one (100\%) and three of three (100\%) in the control and treatment group respectively successfully completed the day's tasks involving outgrowth and pelleting of the culture for analysis by mass spectrometry.
Overall, on first attempt, only one of the five (20\%) control group participants completed all three days whereas three of the five (60\%) participants in the treatment group completed the full experiment on the first attempt.  

When the four participants in the control group that failed on one of the days in the first attempt were allowed to try the experiment again a second time, one additional participant successfully completed the experiment.
Correspondingly, in the treatment group, one of the two participants that failed was able to complete the experiment on the second attempt.
Overall, given two attempts, three of five (60\%) control group participants and four of five (80\%) participants in the treatment group were able to complete the task.

\subsection{Qualitative Observations}
Initial work in designing the study we asked volunteers to use a multimodal version of ChatGPT that allowed for video, audio, and text inputs and outputs. 
Our thinking at the time was that untrained people might respond better to video and audio instructions rather than text.
However, this was not the case as people generally reported that the multimodal user interface was distracting and generally ended up relying on text instructions.
Further, we noticed that the multimodal ChatGPT seemed to struggle in providing instructions on how to use equipment, which we called the `red-button' problem as multimodal ChatGPT seemd to frequently hallucinated that lab equipment had a red button that needed to be pushed. 
In the final version of the study that we presented here, using ChatGPT o1 with deep research enabled seemed to resolve those issues.
Nevertheless, uses did report that the o1's responses were often overly detailed and actually slowed down the process of completing the experiments due to having to sift through lengthy responses.

\section{Discussion} 
The interpretation of our study results should be made with the understanding that this is a pilot study to determine if real uplift is reasonably measurable and how we might conduct such a study. 
Our conclusion is that empirical measurement of real uplift is possible and that empirical studies of the human-AI interface are essential for both AI developers and understanding the threats and opportunities poised by AI for national security.
The results of the study are consistent with skills-based uplift occurring but future work would need more participants to measure both the magnitude and mechanisms of uplift. 
We discuss further some of our thoughts in the design of future uplift studies in real laboratory settings.

\subsection{Future Studies of Real Uplift}

\subsubsection{Protocol Selection Considerations}
Our experience is consistent with the idea that the main technical barriers to measuring real uplift risk fall into the category of what protocol to use to assess uplift.
We argue that a good protocol needs to be (in order of importance) safe, resolving, representative, segmented, and objective. 

\paragraph{Safety} Most important is that the protocol is safe for everyone involved including participants and observers.
Because measuring real uplift risk, by definition, involves people that have little or no experience in a biology laboratory, the protocol must be immediately safe in that it does not use any hazardous materials directly and that no hazardous materials can be formed by any probable failure state of following the protocol.
For example, a protocol that uses harmless materials that, when combined, pose a hazard is not safe even if the protocol explicitly does not involve mixing those compounds.
For similar reasons, study subjects should be observed at all times both for data collection purposes and to avoid safety issues.
If this type of real uplift risk measurement becomes a part of a broader AI risk assessment capacity, it is also a reasonable measure to obtain formal institutional review of protocols that involve human subjects.
While formal review of human-subjects research may not be required for research that is not federally funded, we believe that seeking formal institutional review is an important step in establishing the legitimacy research at the AI-human interface.

\paragraph{Resolution} A protocol with a high resolution is one that very few to zero individuals in the control group (e.g. without access to an AI system) will be able to complete.
The reason for this is that uplift is measured by comparing various outcomes such as successful completion of the protocol in the control versus treatment group; if most people in the control group can complete the protocol, then any experiment based on that procedure will, likewise, not be able to find uplift even if it is occurring.
The issue of low resolution is exacerbated by the fact that studies of real uplift risk will most likely be too small to obtain any real statistical power.
To avoid raising the reader's blood pressure, we argue below that studies of real uplift risk cannot rely on traditional notions of statistical significance and must integrate quantitative outcomes and qualitative observations into broader notion of evidence for or against real uplift risk from the AI system in question. 

\paragraph{Representativeness} Any protocol used in studies of real uplift risk needs to be representative of a procedure that would be used under some reasonable threat model. 
The details of what counts as representative will vary according to the threat model of interest.
For example, a threat model concerned with a lone, unsophisticated actor that requires study participants to use expensive software or high-end computing resources will simply not be representative of protocols consistent with the threat model.
Importantly, we observed that a common point where the multimodal AI system gave poor guidance was in reference to specific equipment in the laboratory, for example, confidently instructing our volunteers to to `press the red button' on a device that has no red buttons.
Given that different threat models will imply very different access to equipment, the specific equipment used in the study should be selected carefully to be representative of the assumptions of the threat model.
The overall representativeness of any real uplift experiment should be taken into account when trying to generalize its results to a specific risk mitigation scenario. 

\paragraph{Segmentation} Given the complexity of measuring real uplift risk, we argue that studies should use a `segmented' structure where a protocol can be split into independent segments that can completed by a study participant even if they failed to complete the proceeding segment.
This type of design has the dual advantages of isolating failure points to a specific aspect of the protocol (e.g. sample preparation versus sample purification), which is useful for understanding how real uplift risk works mechanistically, and ensuring that each study participant attempts each segment of the protocol despite the (ideal) study having high resolving power.

\paragraph{Objectivity} 
By `objectivity' we simply mean that both quantitative and qualitative outcomes are minimally dependent on the subject judgment of an observer.
For example, a study outcome that scores study participants as having produced `a high yield' based on observer notes is not as good as  measuring the optical density of a reaction, enzymatic activity of a sample, or time to complete a task. 
This applies as well for qualitative observations of study participants.
Avoiding inferring what study participants are thinking or feeling and focus on documenting how they are interacting with the AI system itself.
For example, recording which medium study participants used to interact with the AI for multimodal AI systems, and, if possible, what audio, visual, or textual prompts they used. 

\paragraph{Recommendations} We found that the proinsulin protocol that we were prototyping had low resolving power (i.e. volunteers without access to an AI system were able to complete the protocol mostly successfully).
However, as we decided to help volunteers that were stuck or confused, this should not be taken as a ground truth.
Generally, we hypothesize that finding a protocol that is both resolving and representative would require an immense investment in running volunteers though candidate protocols.
Rather than searching for a ideal protocol, we recommend focusing on using a protocol that is consistent with the explicit risk model used in the study and then using additional constraints such a limited time, incomplete reagents, or mislabeled samples to induce a barrier that might be hard to diagnose for the control group (i.e. in the absence of an AI system).

\subsubsection{Outcome Measures}
The notion of uplift is not well defined and therefore difficult to operationalize as a measurement.
We identified several aspects of a reasonable approach to selecting outcome measures for uplift studies as part of our protocol development exercise.

\paragraph{Explicit threat model}
In the context of a explicit threat model (e.g. a narrative context describing the actor, their motivations, and resources), uplift is more clear as a concept. 
For example, assuming a low-skill, lone actor with limited resources seeking to cause a public disturbance, we might simply define uplift as the ability to complete the protocol without access to new equipment and limited, suboptimal reagents.
This could mean simply mean a binary outcome indicating success or failure after some given time.
Alternatively, an actor looking to target specific individuals might need to generate a minimal quantity of a purified toxin, which would be better represented by a measurement of the yield of a proxy (non-toxic) compound.
The overall notion of what uplift is should flow naturally from an explicitly defined, but abstract, risk model.

\paragraph{Description over power}
After attempting to run volunteers though the test protocol to see where problems would arise, we concluded that seeking out statistical power or any notion of formal hypothesis testing does not make sense for uplift studies.
The context is simply too variable (e.g. heterogeneity in aptitude, psychological disposition, or motivation) and the outcome too difficult to measure to rely on a formal hypothesis testing framework.
Rather, we argue that the goal of a well-designed study of uplift should be a rich dataset that makes multiple, well defined measurements of success at key points along the (ideally) modular protocol.
For example, at each segment of the protocol, record the the number and type of mistakes such as inappropriate use of reagents in addition to a simple measure of success such as completion of the protocol segment in the allotted time.  
This approach generates a much richer description of how the AI system changes how study participants interact with the tools, which we argue should be the goal of a well-designed uplift study.
The context in which studies of uplift will be used for regulatory or design purposes is not currently a `go/no-go' situation where we need some unambiguous single answer. 
Outcome measures should reflect the goal of better understanding how AI systems are actually changing peoples' behavior rather than trying to statistically test if uplift occurred as a single binary inference.
For example, we noticed that volunteers in the AI group encountered AI hallucinations concerning the specific details of the laboratory equipment, creating substantial confusion in the volunteers.
Understanding how the use of AI might change how novices react to errors could lead to novel risk mitigations such as refusal to give specific instructions for operating equipment common to biological laboratories. 

\paragraph{Anthropological observations are a primary outcome}
We were surprised by how informative observations of how volunteers were interacting with the multimodal AI system were to our overall insights.
For example, some volunteers expressed being distracted by the AI interjecting questions too frequently or non-specific annoyance at the voice the AI used to talk to the volunteers or that setting up the camera to get a good view of their workspace was difficult. 
Some of these problems were exacerbated by our use of multimodal AI that included video, voice, and text input/output, which relies on user interfaces that are not yet as refined as public-facing frontier model interfaces.
Further, we did not provide specific guidance beyond general instructions to our volunteers on the specific use of the multimodal AI system. 
However, we believe that the vision, video, and voice capabilities of multimodal AI systems will likely be essential to a `lab assistant' type of AI that would be most useful in a broad range of risk models based on low-skill actors.

Fundamentally, risk generated by AI systems comes from some capability of the AI system itself that is itself value neutral and only becomes risky through the intentions and actions of an actor.
Therefore, we believe that it is essential to observe and document narratives of how study subjects actually use the AI system under study to understand both how risk might be mitigated but also how beneficent use can be encouraged.
These anthropological observations should not be considered a secondary concern but rather a co-equal outcome of the study along whit any quantitative data generated.
The motivation for this recommendation, beyond the  descriptive power gained by combining qualitative and quantitative data into a wholistic accounting of the mechanisms of uplift risk, is more clear in the context of the goal of these studies.

\subsubsection{Study Design}
We identified several issues that fall into the general category of study design.

\paragraph{Isolation and Observation}
The type of threat models that we were thinking about when we started this work involved lone or mostly isolated individuals that are effectively working alone.
From that perspective, we believed that it was important to isolate study subjects from one another as much as is possible to prevent subjects from copying each other.
However, understanding how AI use might lead to real uplift in a group setting is also important from the perspective of a threat model that is concerned with real uplift at the group or institutional levels.
There is no single right answer, but the choices should correspond to aspects of the threat model of interest.

Regardless of the design choices in studies of real uplift, study participants need to be directly observed, both for data collection and safety, while they are attempting to complete the protocol. 
Ideally multiple observers would record observations independently to avoid perceptual biasses and to minimize the probably of missing important observations.
Observers should have a clear standard for how they should interact with study subjects, such as when to intervene for safety concerns and how to avoid contaminating the study results by interacting with the study participants. 

\paragraph{Sample Size}
Traditional elements of experimental study design such as statistical power are not as rule-bound in this type of descriptive study as they might be in a study focused on hypothesis testing.
We suspect that a reasonable rule-of-thumb is that the size of the treatment group should be about as large as the inverse of the probability that a hypothetical threat actor would stop attempting to use the AI system given a recent failed attempt.
For example, for a study assuming a highly dedicated actor might only have a 1\% chance of giving up if they failed recently, a treatment group size of at least 100 might be a reasonable starting point under the logic that you want to detect successful uplift that occurs about 1\% of the time.
However, just because there is no objective way to determine an `optimal' study size does not mean that smaller studies are sufficient; all else being equal, a large study is more likely to capture important human-AI interactions and is, therefore, more powerful.

\paragraph{Study population and Compensation}
Drawing study participants from a truly representative population for this kind of study is probably impossible.
However, the unique qualities of populations such as graduate and undergraduate students that are accessible though university programs and, most likely, should be described and explicitly noted. 
The objective is not to identify that uplift occurs in one population but not another.
Rather, we want to build up a scientific understanding of how uplift works mechanistically and how different people interact with AI systems.
For example, if younger people interact with AI systems fundamentally differently than older people, we want to understand how those differences change the potential for uplift.
From that perspective heterogeneity in the study group is desirable, however, to obtain sufficient observations in a highly heterogeneous group requires much larger sample sizes that might not be feasible.
The key element is that individual-level demographics of the study participants be recorded as much as is allowable to be compliant with human-subjects institutional review.

Study subjects could be compensated to emulate the motivation that a malicious actor might have. 
However, compensation could attract individuals who are motivated by the compensation and have a skewed approach to the experimental context.
Generally, we assume that most real uplift studies will be performed with populations of college students who are already likely to possess high levels of motivation and baseline knowledge of AI systems.
Regardless, if compensation is provided it should follow basic ethical standards that it should not be an amount that is coercive to study participants or inspires a feeling of indebtedness that could alter the study results. 

\subsubsection{Future Direction}
Empirical measurement of uplift risk is expensive and time-consuming: designing and calibrating protocols, obtaining human-subject review, running experiments, and interpreting results could take weeks to months.
This fact alone places a hard limitation on the role that empirical measurement of real uplift risk can have in an ongoing assessment of AI risk.
Running these kinds of protocols is neither viable for every new release of an AI system, nor as a single baseline measurement of AI systems generally; AI systems change very rapidly and even changes in the user interface could enhance or inhibit real uplift.
Given the capabilities of AI systems are rapidly evolving and our understanding of how the capabilities those AI systems bring to non-experts induce risks to populations and systems is limited, we recommend a hybrid approach focused around extracting scalable benchmarks from rigorous, large-scale studies of real uplift.
That is, running one or many large-scale real uplift experiments, possibly through a consortium effort, and identifying exactly where and how uplift is occurring and then making a scalable benchmark that can be simulated (e.g. creating video, text, and audio) and posed to current and future AI systems at scale.

In the process of running these protocol development exercises, we already identified the interaction between AI, humans, and equipment to be a likely point of failure in real uplift, and also a potential locus of control for mitigating uplift risk.
Using experiments to identify how real people interact with AI systems in the context of an uplift task, we can avoid biases or omissions that might prevent us from seeing how uplift actually works in non-experts.
Sharing the findings from future uplift studies will be useful for not only mitigating risk but also enabling beneficial use of AI systems for legitimate discovery.

\section{Funding and Conflicts of Interest}
Research presented in this report was supported by the Laboratory Directed Research and Development program of Los Alamos National Laboratory under project number 20250639DI and through Los Alamos National Laboratory as part of the National Security and International Studies Fellowship for EORS.
Access to ChatGPT (multimodal and o1) was provided by OpenAI. 
We have no other financial, personal, or professional conflicts of interest to declare.

\section{Specific Contributions}
Conception: NG, PM; Execution: TH and PM; Analysis: EORS, TH, PM, NG; Writing: EORS; Revisions: EORS, PM, TH, NG; Funding: NG; Administration: NG, PM, TH, EORS

\section{Acknowledgments}
We would like to thank Miles Wang and Tejal Patwardhan at OpenAI for their technical assistance and fruitful discussions. 



\smallskip

\setstretch{1}
\bibliography{refs}
\bibliographystyle{compactunsrt}
\setstretch{\stretchby}

\end{document}